\documentclass[aps,prb,twocolumn,showpacs,superscriptaddress,10pt]{revtex4-1} 

\usepackage{graphicx}
\usepackage[utf8]{inputenc}
\usepackage{parskip}
\usepackage{xcolor}
\usepackage{amssymb}
\usepackage{amsmath}
\usepackage{mathtools}
\usepackage{physics}
\usepackage{bm} 
\usepackage[percent]{overpic}
\usepackage{microtype}
\usepackage[caption=false]{subfig}
\usepackage{makecell}
\usepackage{siunitx}
\usepackage[pdftex]{hyperref}

\newcommand{\ie}{\emph{i.e.}}

\usepackage[]{nth} 
\usepackage[colorinlistoftodos,prependcaption,textsize=scriptsize]{todonotes} %

\newcommand\unit[1]{\,\mathrm{#1}}

\newcommand\pa[1]{\left ( #1 \right )}
\renewcommand\pb[1]{\left [ #1 \right ]}

\begin{document}
\title{Edge-dependent reflection and inherited fine structure of higher-order plasmons in graphene nanoribbons}

\author{K\aa re Obel Wedel}
\affiliation{Department of Photonics Engineering, Technical University of Denmark, {\O}rsteds Plads, Bldg.~345A, DK-2800 Kongens Lyngby, Denmark}
\affiliation{Department of Physics, Technical University of Denmark, Fysikvej, Bldg.~307, DK-2800 Kongens Lyngby, Denmark}
\affiliation{Center for Nanostructured Graphene (CNG), Technical University of Denmark, {\O}rsteds Plads, Bldg.~345C, DK-2800 Kongens Lyngby, Denmark}
\author{N. Asger Mortensen}
\affiliation{Center for Nano Optics, University of Southern Denmark, Campusvej 55, DK-5230 Odense M, Denmark}
\affiliation{Danish Institute for Advanced Study, University of Southern Denmark, Campusvej 55, DK-5230 Odense M, Denmark}
\affiliation{Center for Nanostructured Graphene (CNG), Technical University of Denmark, {\O}rsteds Plads, Bldg.~345C, DK-2800 Kongens Lyngby, Denmark}

\author{Kristian S. Thygesen}
\affiliation{CAMD, Department of Physics, Technical University of Denmark, Fysikvej, Bldg.~307, DK-2800 Kongens Lyngby, Denmark}
\affiliation{Center for Nanostructured Graphene (CNG), Technical University of Denmark, {\O}rsteds Plads, Bldg.~345C, DK-2800 Kongens Lyngby, Denmark}

\author{Martijn Wubs}
\affiliation{Department of Photonics Engineering, Technical University of Denmark, {\O}rsteds Plads, Bldg.~345A, DK-2800 Kongens Lyngby, Denmark}
\affiliation{Center for Nanostructured Graphene (CNG), Technical University of Denmark, {\O}rsteds Plads, Bldg.~345C, DK-2800 Kongens Lyngby, Denmark}

\begin{abstract}
We investigate higher-order plasmons in graphene nanoribbons,  and present how electronic edge states and wavefunction fine structure influence the graphene plasmons. Based on nearest-neighbor tight-binding calculations, we find that a standing-wave model based on nonlocal bulk plasmon dispersion is surprisingly accurate for armchair ribbons of widths even down to a few nanometers, and we determine the corresponding phase shift upon edge reflection and an effective ribbon width.
Wider zigzag ribbons exhibit a similar phase shift, whereas the standing-wave model describes few-nanometer zigzag ribbons less satisfactorily, to a large extent because of their edge states. We directly confirm that also the larger broadening of plasmons for  zigzag ribbons is due to their edge states.
Furthermore, we report a prominent fine structure in the induced charges of the ribbon plasmons, which for armchair ribbons follows the electronic wavefunction oscillations induced by inter-valley coupling.
Interestingly, the wavefunction fine structure is also found in our analogous density-functional theory calculations, and both these and tight-binding numerical calculations are explained quite well with analytical Dirac theory for graphene ribbons.
\end{abstract}

\date{\today}

\maketitle

\section{Introduction}

Numerous studies have over the recent years been conducted on graphene one-dimensional (1D) structures, emphasizing both single-particle excitations and collective plasmonic excitations.\cite{BrarHighlyNanoresonators,Low2014GrapheneApplications,Stauber2014PlasmonicsInsulators,Fei2012Gate-tuningNano-imaging,Ju2011GrapheneMetamaterials,Yan2013DampingNanostructures,Fei2015EdgeNanoribbons,Xu2016EffectsNanoimaging}
Ribbons are prime examples of such structures,\cite{Christensen2012GrapheneNanoribbons,Thongrattanasiri2012QuantumInformation,Silveiro2015QuantumNanoribbons} while plasmons can also be localized and guided along other 1D structures.\cite{Goncalves:2017a,Goncalves:2017b,Goncalves:2017c}
Principal motivations for studying plasmons in graphene ribbons are the strong confinement of the electromagnetic fields, long propagation lengths, as well as the convenient tunability through (electrostatic) doping.\cite{Huang2016GrapheneApplications}

Creation of nanoribbons has come a long way.\cite{Segawa2016StructurallyNnostructures,Koga2018SynthesisByDimerization,Cai2010AtomicallyNanoribbons,Narita2014SynthesisNanoribbons,Narita2015NewChemistry,Ruffieux2016On-surfaceTopology,Wang2016GiantEdges}
It is now possible to create ribbons in the 10--20 nm range both with top-down processes, allowing better scalability, and with bottom-up syntheses yielding high atomic precision.\cite{Xu2016RecentFabrication}
Together with methods for probing plasmons with high spatial resolution\cite{Chen2012OpticalPlasmons,Fei2012Gate-tuningNano-imaging,Fei2015EdgeNanoribbons,Woessner2014HighlyHeterostructures,Low2016PolaritonsIn2D} this creates possibilities to measure novel quantum effects in graphene plasmonics.

\begin{figure}[b]
\centering
\includegraphics[width=3.4in]{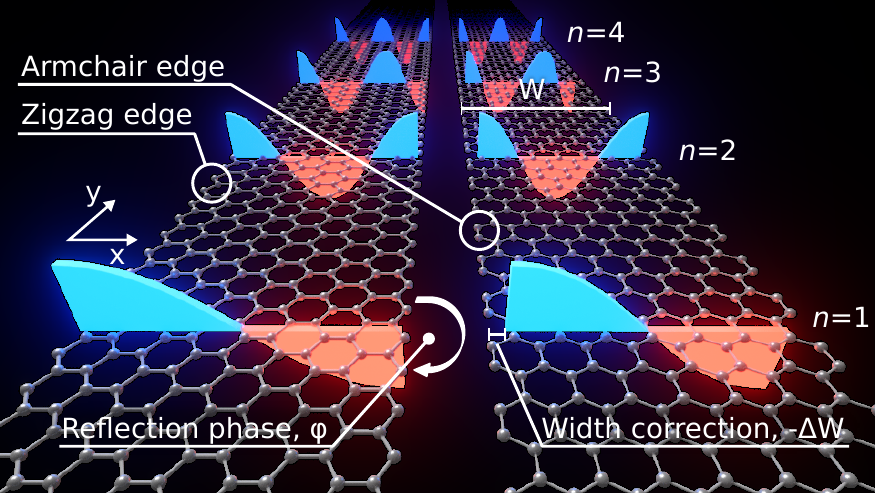}
\caption{(Color online) A zigzag (left) and an armchair (right) ribbon with the axis used in the following being indicated. Induced charges across the ribbon for dipolar and higher-order plasmons are illustrated in red (negative charges) and blue (positive). }\label{fig:introimage}
\end{figure}

We have previously elucidated the emergence of nonclassical behavior of the lowest-order plasmons in narrow graphene ribbons~\cite{Wedel2018} arising from the quantized nature of the bands.
In this work, we analyze instead the higher-order modes, in order to study the impact of the precise atomic configuration on the plasmon reflection properties of the ribbon edges.
The phase shift upon edge reflections of plasmons in graphene has previously only been treated in continuum theories, in Refs.~\onlinecite{BrarHighlyNanoresonators,Nikitin2014AnomalousResonators,Velizhanin2015GeometricArrays,Christensen2015FromDimensions}, where conductivity is handled as a local material parameter.
Possible effects of the specific atomic configuration at the edge cannot be studied in such an analysis.
In contrast, we here study edge reflections within tight-binding (TB) calculations for both armchair and zigzag ribbons (see Fig.~\ref{fig:introimage}).
We also consider zigzag ribbons where the edge states have been excluded when calculating the optical response as detailed in our previous work.\cite{Wedel2018}
The latter allows us to study directly how  graphene plasmons are affected by the localized electronic edge states of zigzag ribbons.

Furthermore, the atomistic nature of our calculations allows us to study the fine structure of the plasmons by mapping the induced charges to individual atomic sites.
The analysis reveals short-range oscillations inherited from the underlying wavefunctions, predicted by Dirac theory and confirmed both by TB and our \emph{ab initio} density-functional theory (DFT) calculations. 

The structure of the paper is as follows:
In Sec.~\ref{sec:standingwaves} we present our analysis of a standing-wave model and the effect of the atomic edge termination on the edge reflection properties of graphene plasmons.
Secondly, in Sec.~\ref{sec:edgestatesinducedbroadening}, we briefly show our findings regarding the localized edge states' ability to introduce additional broadening of the plasmonic peaks.
Lastly, we dive into the spatial distributions of the plasmons and the differences in the induced fine-structure in Sec.~\ref{sec:finestructure}.

\section{Models and methods}\label{sec:methods}

\subsection{Tight-binding model}\label{sec:TB}
The band structure of graphene is well described by a  nearest-neighbor TB model  with the Hamiltonian
\begin{align}
H = \sum_{<i,j>} -t(a^\dagger_ib_j + h.c.),
\end{align}
where the sum is over pairs of neighboring sites.\cite{CastroNeto2009}
For the hopping parameter $t$ we use the value of 2.8\,eV, first determined by Ref.~\onlinecite{Wallace1947TheGraphite}.

The eigenstates are calculated on a dense $k$-point grid with 5000 points in the one dimensional Brillouin zone and used for calculating the optical response as outlined below.
In ribbons with zigzag edges (left ribbon in Fig.~\ref{fig:introimage}) where localized edge states occur, we can classify the eigenstates as either bulk-like or edge-like using an energy cutoff derived from the Dirac model as presented in our recent work (Ref.~\onlinecite{Wedel2018}).
This will allow us to directly quantify the effect of the edge states on the energies and reflection properties of the graphene plasmons.

\subsection{Response function}\label{Sec:Response_function}

We calculate the optical response for $q=0$ within the random-phase approximation (RPA) following the same methodology as Refs.~\onlinecite{Thongrattanasiri2012QuantumInformation,Wedel2018}, \ie\ the non-interaction density-density response function is calculated in the site basis through direct insertion of the eigenstates in\cite{Bruus:2004}
\begin{align}
\chi_{ij}^0(\omega) = \frac{2e^2}{\hbar} \frac{b}{2\pi}\int\limits^{\mathrm{BZ}}\!\!{\mathrm{d}k} \sum_{nm} f_{nm}\frac{a^{}_{in}a^*_{im}a^*_{jn}a^{}_{jm}}{\epsilon_{nm}+\hbar(\omega + i\eta)}, \label{eq:suscep_def}
\end{align}
 from which the dielectric function can be determined as
\begin{align}
\epsilon_{ij} = 1 - V_{il}\chi^0_{lj},\label{eq:epsilon_rpa_def}
\end{align}
where $V$ is the Coulomb interaction. The $i,j$ are atomic site indices, while $n$ and $m$ label the eigenmodes at wave vector $k$.
Thus, $a_{in}$ is the value of the $n^\mathrm{th}$ wavefunction on the $i^\mathrm{th}$ site (implicitly at wave vector $k$).
As a shorthand notation, we have used $\epsilon_{nm} = \epsilon_n - \epsilon_m$ for the energy difference and likewise $f_{nm} = f_n - f_m$ for the difference in the Fermi filling factors.
The phenomenological loss parameter $\eta$ is set to $1.6\unit{meV}$ as in Ref.~\onlinecite{Thongrattanasiri2012QuantumInformation}.
The width of the supercell in the periodic direction is labeled $b$.
By excluding the edge states in the evaluation of the response function, their contribution can be assessed by comparing with the full expression.

The Coulomb interaction is included in real space using tabulated values for the correct interaction between $p_z$ states.\footnote{Same data as in Ref.~\onlinecite{Thongrattanasiri2012QuantumInformation} and acquired through private correspondence with the group.} Charge neutrality ensures that the  product $V\chi^0$ can be properly converged, despite the long-range behavior of the Coulomb interaction.\cite{Thongrattanasiri2012QuantumInformation,Wedel2018}

\subsection{Quantum plasmons}

\begin{figure}[htbp]
\centering
\includegraphics[width=3.4in]{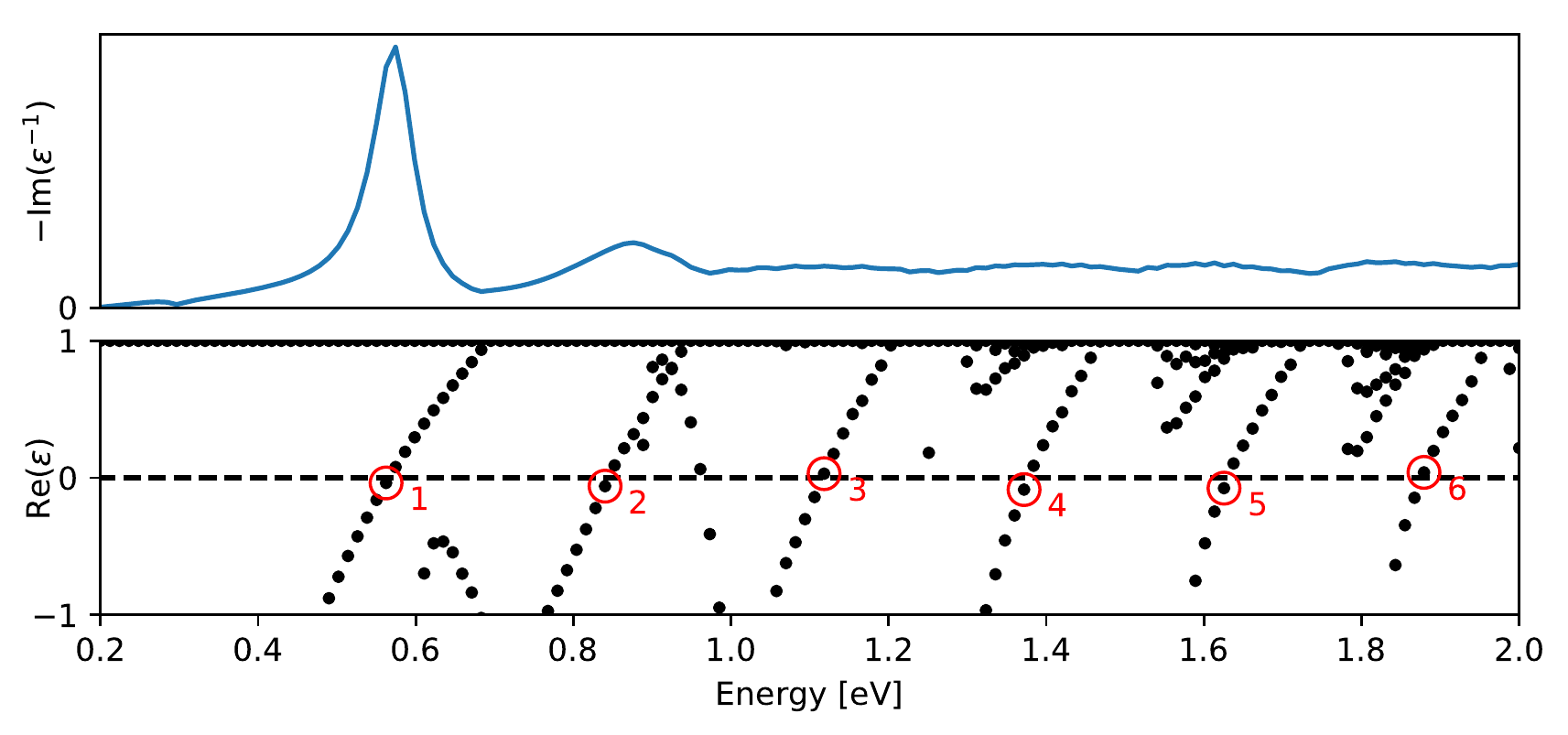}
\caption{(Color online) From the dielectric matrix the plasmon modes can be found as peaks in the loss function (top panel) where the dipole plasmon stands out, or as the zeros of the real part of $\epsilon_n$ as shown in the lower panel. The data shown is for a $6\unit{nm}$ zigzag ribbon with $\epsilon_F = 0.4\unit{eV}$.}\label{fig:eps_n}
\end{figure}

The dielectric function $\epsilon(\omega)$ can be written in a spectral representation of its eigenvalues and left and right eigenvectors as
$\epsilon_{ij}(\omega) = \sum_n \epsilon_n(\omega)\phi_{n,i}(\omega)\rho^*_{n,j}(\omega)$,
where the zeros of the real parts of $\epsilon_n(\omega)$ indicates plasmonic modes, the right eigenvector $\phi_n$ is the induced field, and the left eigenvector $\rho_n$ is the induced charges of the plasmon.~\cite{Andersen2012SpatiallyFirst-principles}
In Fig.~\ref{fig:eps_n} the numerically calculated eigenvalues for a $6\unit{nm}$ wide ribbon with zigzag termination and a Fermi energy of $0.4\unit{eV}$ are shown below the panel showing the energy loss function, the latter defined as $-\Im(\epsilon^{-1})$.
The crossings of zero by the real part of the eigenvalues are indicated with red circles.
The first two zeros of $\Re[\epsilon_n]$ clearly correspond to   peaks in the loss spectra. Higher-order modes are more  damped and hard to identify from the loss spectrum, but they can still be easily identified as the zeros of $\Re[\epsilon_n(\omega)]$. 

\section{Standing-wave model}\label{sec:standingwaves}

It is well known that plasmons reflect with almost no loss on graphene edges.\cite{Chen2013StrongPlasmonReflection,GarciaPomar2013ScatteringGPlasmons}
Thus, as a method of understanding the behavior of plasmons in graphene nanoribbons, we will adopt a Fabry--P{\'e}rot standing-wave model.
As we only consider propagation in the $x$ direction, the picture is that the plasmon moves across the ribbon according to a certain dispersion relation, reaches an edge, and reflects back with an additional phase change from the reflection.
The allowed modes are those where this process gives rise to constructive interference as illustrated in Fig.~\ref{fig:introimage}.
The condition for this to occur becomes
\begin{equation}
2(n-1)\pi = 2qW_\mathrm{eff} + 2\varphi \Leftrightarrow
q = \frac{(n-1)\pi - \varphi}{W + \Delta W},
\end{equation}
where $n$ is the integer mode index starting from $n=1$ and $\varphi$ is the reflection phase change. Furthermore we introduced an effective width $W_\mathrm{eff} \equiv W + \Delta W$ that takes into account that the plasmon may not reflect at exactly the positions of the outermost rows of atoms that define the geometric width $W$.
The notion of effective sizes are also found in the area of optical antennas.\cite{Novotny2007EffectiveScaling}
A positive $\Delta W$ describes a plasmon that effectively spills out of the ribbon, while a negative value corresponds to a plasmon that is effectively more tightly confined than by the geometric width. As such, this is quite analogous to descriptions surface phenomena based on Feibelman parameters.\cite{Feibelman:1982a,Christensen:2017}

We have performed TB calculations for both armchair and zigzag ribbons and also considered zigzag ribbons where the edge states have been excluded when calculating the optical response, as detailed in our previous work.\cite{Wedel2018} 
This allows us to understand the effects, if any,  of the atomic edge termination and the localized edge states on the reflection properties of the graphene plasmons.

\subsection{Linear mode dependence of higher-order modes}\label{sec:linearhighermodes}

\begin{figure*}[htbp]
\centering
\includegraphics[width=7in]{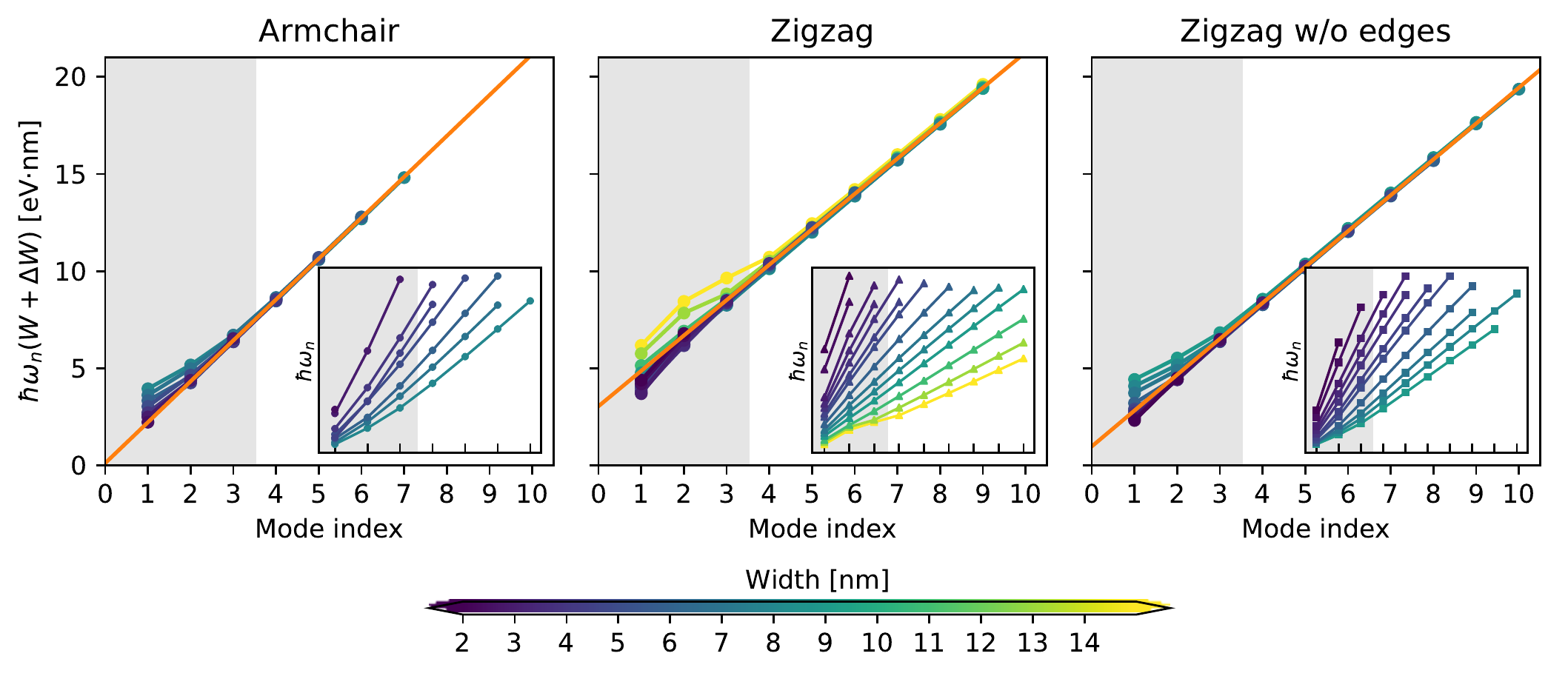}
\caption{(Color online) Using a linear dispersion relation and fitting the Fabry--P{\' e}rot model to the modes with $n \geq 4$ for AC, ZZ, and ZZ without edge states. The insets show the energy as a function of mode number for all the ribbons calculated. All calculations are for $\epsilon_F =0.4\unit{eV}$.}\label{fig:mode_dependency}
\end{figure*}

By finding the zeros of the real part of the eigenvalues of the dielectric matrix, as illustrated in the bottom panel of Fig.~\ref{fig:eps_n}, we can find the plasmon energies as a function of mode index.
We depict this data in the insets of Fig.~\ref{fig:mode_dependency}.
By inspection one can see that the plasmon energies depend more or less linearly on the mode number for the higher-order modes.
Given this linear dependence, it seems that the higher-order plasmons on graphene ribbons behave analogously to light in a cavity between two mirrors. Assuming a linear dispersion as $\omega_n = v_\mathrm{p} q_n$, where $v_\mathrm{p}$ is a constant plasmon velocity, we therefore expect $\omega_n W_\mathrm{eff}$ to be constant across different widths.
To fit our non-dispersive model we do not use the lowest-order modes with $n\leq 3$, as indicated by the gray areas in Fig.~\ref{fig:mode_dependency}. The reason is that the curves shown in the insets start deviating  from the linear behavior for these lower mode numbers.
The resulting fits are shown in Fig.~\ref{fig:mode_dependency} and the corresponding values are given in Tab.~\ref{tab:linear_fit}.
\begin{table}[htbp]\renewcommand{\arraystretch}{1.8}
\centering
\caption{Fitting parameters as determined from the linear dispersion model used in Fig.~\ref{fig:mode_dependency}.}\label{tab:linear_fit}
\sisetup{table-figures-uncertainty=1, separate-uncertainty=true, table-number-alignment=center}
\begin{tabular}{@{}lSSS@{}}
\toprule
{}& {Armchair} & {Zigzag} & {\makecell{Zigzag w/o\\edge states}} \\ \colrule
$\Delta W$ [nm] & 0.38 \pm 0.05 & 1.44 \pm 0.04 & 0.72 \pm 0.02 \\
$\varphi/\pi$ & -1.06 \pm 0.05 & -2.67 \pm 0.05 & -1.53 \pm 0.03\\
$v_\mathrm{p}$ [$10^6\unit{m/s}$] & 1.02 \pm 0.02 & 0.88 \pm 0.00 & 0.90 \pm 0.00 \\
\botrule
\end{tabular}
\end{table}
The linear fit is indeed quite good for the higher-order modes in all cases.
Without edge-state contributions there is a slight upward bending of the lower-order modes that gets more prominent for the wider ribbons.
When comparing ZZ with and without edges, we can tell that the edge states alter the behavior of the low-index modes, while the higher-order modes are still linear.
The extracted plasmon velocities differ by $\sim$10\% and are all close to the Fermi velocity, $v_\mathrm{F} \approx \SI{0.91e6}{\meter\per\second}$.

As seen in Tab.~\ref{tab:linear_fit}, in this model AC edges have a reflection phase of approximately $-\pi$ and a small width correction $\Delta W \simeq 0.4\unit{nm}$.
The zigzag ribbons show a very different behavior with a larger $\Delta W$ of $1.44\unit{nm}$ and a considerable phase shift of $-2.67\pi$.
Removing the edge states brings both $\varphi$ and $\Delta W$ closer to the results found for armchair ribbons.

Although the linear fits are quite good, the model only works for the higher-order modes and the more-than-$2\pi$ phase shift for zigzag ribbons is hard to interpret. We therefore conclude that a better model is needed to obtain trustworthy quantitative values for the $\varphi$ and $\Delta W$.
This model will be presented in the following.

\subsection{Nonlocal dispersion and reflection phase shift}\label{sec:nonlocaldispersion}

\begin{figure*}
\centering
\includegraphics[width=7in]{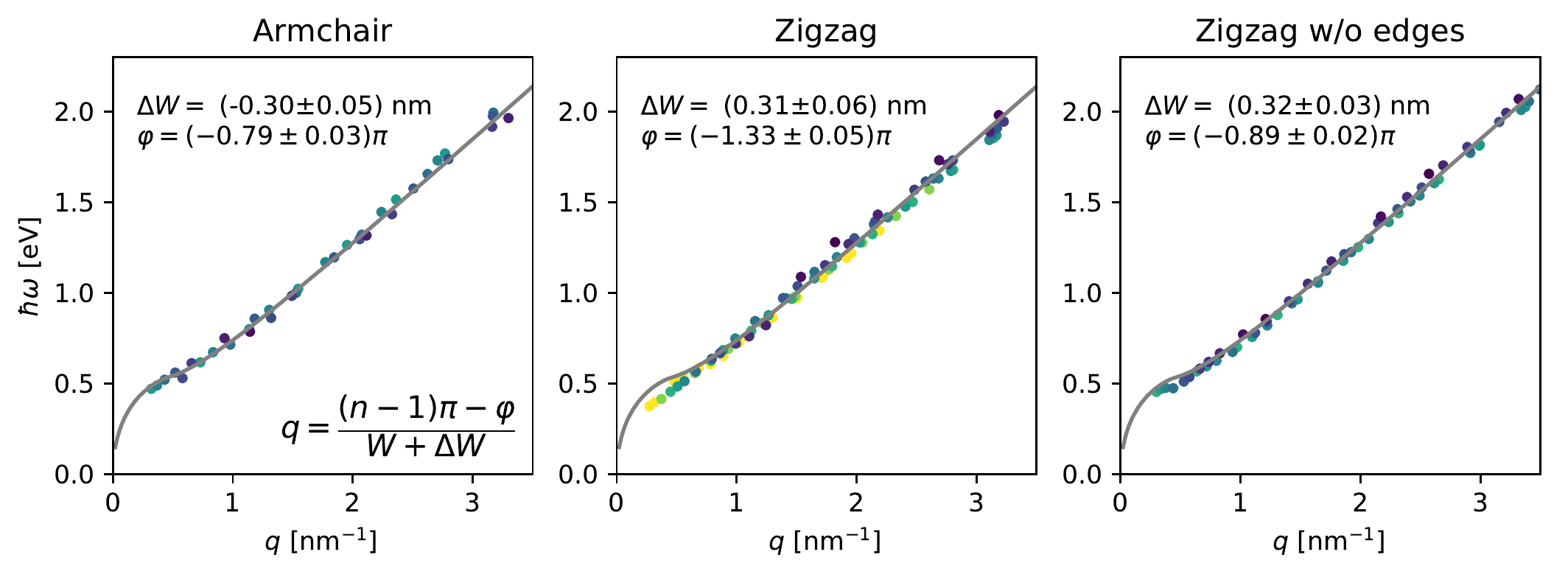}
\caption{(Color online) The reflection phase and the width corrections are found by optimizing to the nonlocal plasmon dispersion of infinite graphene. The Fabry--P{\' e}rot model with this dispersion works very well for the armchair ribbons and for the zigzag ribbons when excluding the edge states.}\label{fig:non_local_fit}
\end{figure*}

Building on the standing-wave model, we suggest that, while the plasmon is not at the edges, it disperses in the same manner as it would in an infinite sheet of graphene.
Classically, that corresponds to a $\sqrt{\smash[b]{q}}$-dispersion, as is the case for the two-dimensional (2D) electron gas.\cite{Hwang2007DielectricGraphene,Santoyo1993PlasmonsDimension}
However, we expect nonlocality to play an important role in these small structures and we thus use the dispersion relation found by using the nonlocal dielectric function for infinite graphene as calculated in Refs.~\onlinecite{Hwang2007DielectricGraphene,Wunsch2006DynamicalDoping}.
With this approach, an explicit $q$-dependence is included in the quantum mechanical conductivity altering the plasmon dispersion for larger values of $q$.
As can be seen from Fig.~\ref{fig:non_local_fit}, the included nonlocality makes the dispersion almost linear at larger $q$ and thus explains why the linear model worked for high mode indices.

We determine $\varphi$ and $\Delta W$ by fitting to the nonlocal dispersion curve getting the results shown in Fig.~\ref{fig:non_local_fit} with parameters shown in Tab.~\ref{tab:non_local_fit}.
The model applies very well for the armchair ribbons, both for larger $q$ values where the dispersion is linear, and for smaller $q$ where the dispersion curve becomes flatter. The resulting plasmon reflection phase for AC ribbons is found to be close to $-0.75\pi$. The concomitant width correction $\Delta W \approx -0.3\unit{nm}$ corresponds approximately to the width of two and a half atomic rows in the armchair configuration.

An alternative definition of the reflection phase (that differs by $\pi$) has been used in Refs.~\onlinecite{BrarHighlyNanoresonators,Velizhanin2015GeometricArrays,Christensen2015FromDimensions}.
However, after converting to our definition these works report reflection phases that are all very close to $-0.75\pi$.
This is the same as was found in Ref.~\onlinecite{Nikitin2014AnomalousResonators} that uses the same definition as we do.
Because of this remarkable agreement in numerically determined reflection phases, it is worth mentioning at this stage that as far as we know there is no analytical theory that predicts an exact reflection phase of $-3\pi/4$. However, in 
Ref.~\onlinecite{Nikitin2014AnomalousResonators} the authors do present an analytical model that comes quite close and predicts $\varphi\approx -0.64\pi$.

The same nonlocal-dispersion model does not agree as accurately with the analogous tight-binding results for zigzag ribbons, as can be seen from the increased scatter of the points in the second panel of Fig.~\ref{fig:non_local_fit} .
Especially the behavior of the low-$q$ plasmons in the TB calculations is not captured that well.
As seen in the rightmost panel, removing the edge states does improve the agreement, indicating that these states are responsible for a great part of the difference with armchair ribbons.
We emphasize that the AC ribbons are well described by a $-0.75\pi$ reflection phase in combination with the bulk plasmon dispersion down to very small sizes of only a few nanometers.
However, because of the less convincing fit for the ZZ geometry, we will not  take the resulting fitting parameters at face value, and perform instead an additional more thorough analysis.
\begin{table}[htbp]\renewcommand{\arraystretch}{1.8}
\centering
\caption{Fitting parameters as determined from the nonlocal dispersion model used in Fig.~\ref{fig:non_local_fit}.}\label{tab:non_local_fit}
\sisetup{table-figures-uncertainty=1, separate-uncertainty=true, table-number-alignment=center}
\begin{tabular}{@{}lSSS@{}}
\toprule
{}& {Armchair} & {Zigzag} & {\makecell{Zigzag w/o\\edge states}} \\ \colrule
$\Delta W$ [nm] & -0.30 \pm 0.05 & 0.31 \pm 0.06 & 0.32 \pm 0.03 \\
$\varphi/\pi$ & -0.79 \pm 0.03 & -1.33 \pm 0.05 & -0.89 \pm 0.02\\
\botrule
\end{tabular}
\end{table}

\subsection{Width-dependent phase shift}\label{sec:withdependentphaseshift}

\begin{figure*}[htbp]
\centering
\includegraphics[width=7in]{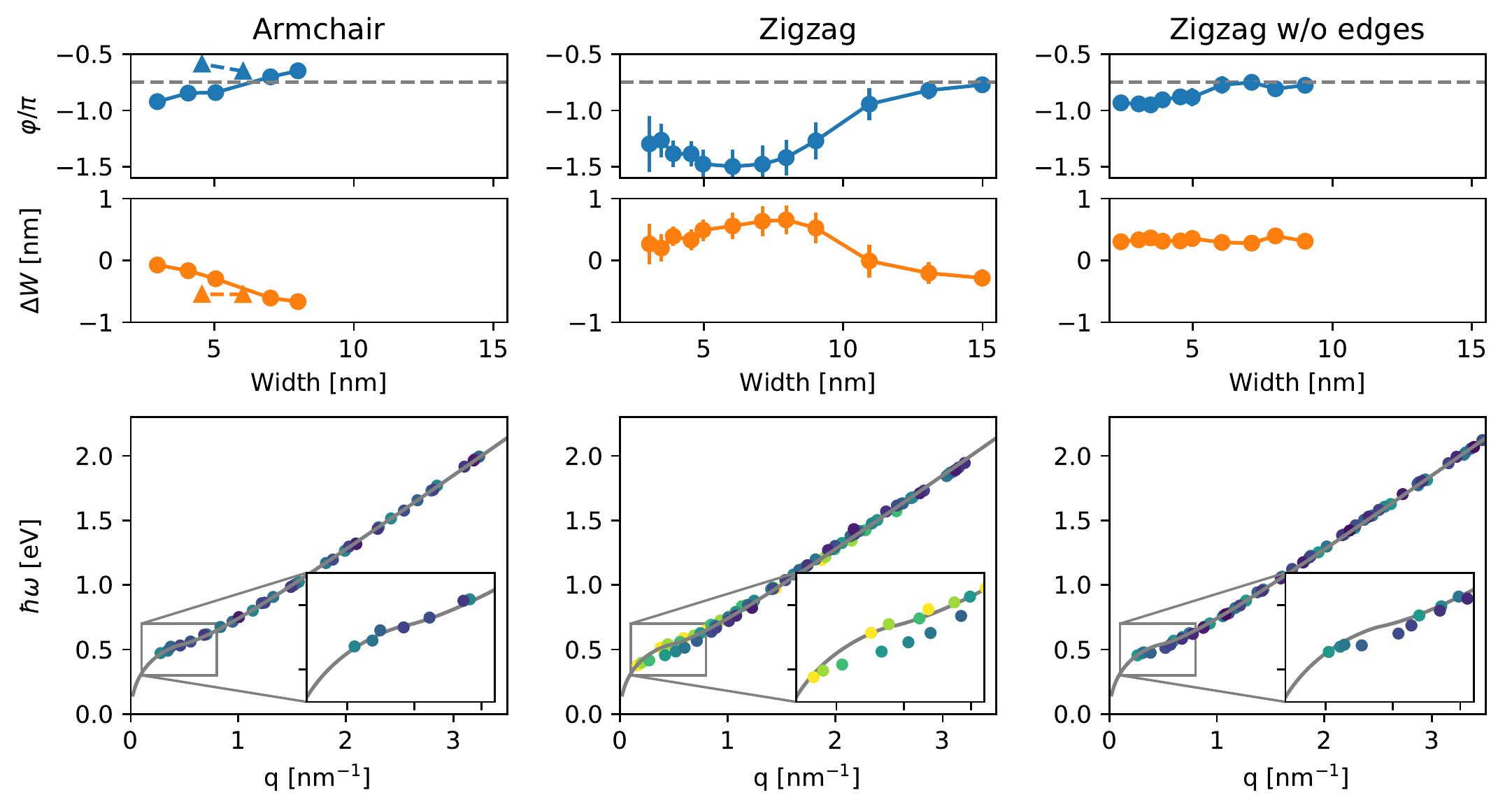}
\caption{(Color online) Optimizing $\varphi$ and $\Delta W$ for one width at a time showing that while the AC results are fairly constant, ZZ corrections seem to converge only for wider ribbons. The two types of points in the AC plots distinguish between semi-metallic (triangles) and semiconducting (circles) ribbons. The dashed line in the top plots indicates $-0.75$. Colors in bottom plots are the same as in Fig.~\ref{fig:mode_dependency}.}\label{fig:width_dep_phase}
\end{figure*}

\begin{figure}[htbp]
\centering
\includegraphics[width=3.4in]{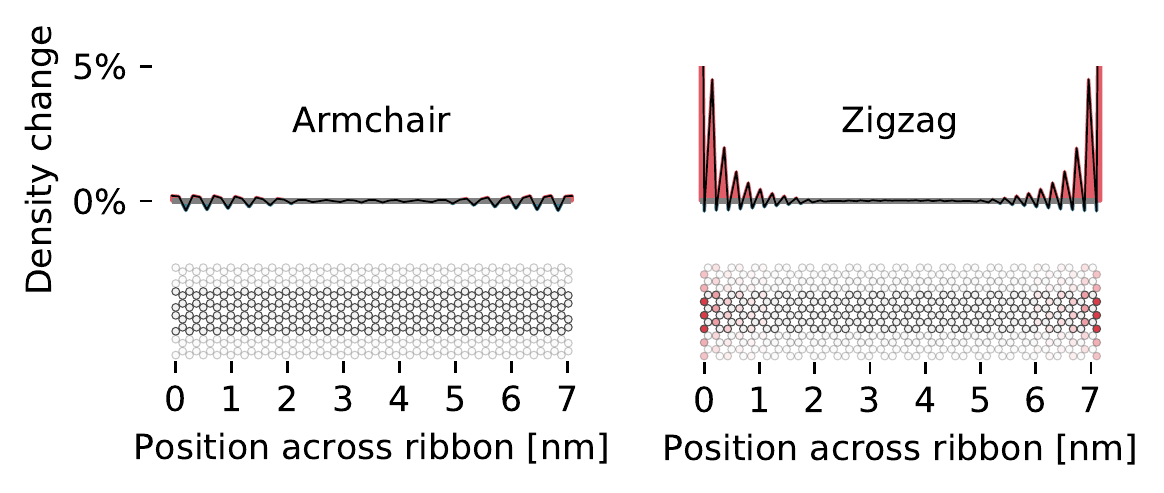}
\caption{(Color online) The difference in the ground-state density for a 7 nm wide doped graphene ribbon, shown relative to the average density at the center half. While the density in AC ribbons is almost constant everywhere, the electronic edge states in ZZ ribbons alter the picture considerably. Results from TB with $\epsilon_F = 0.4\unit{eV}$.} \label{fig:ribbondens}
\end{figure}

\begin{figure}[htbp]
\centering
\includegraphics[width=3.4in]{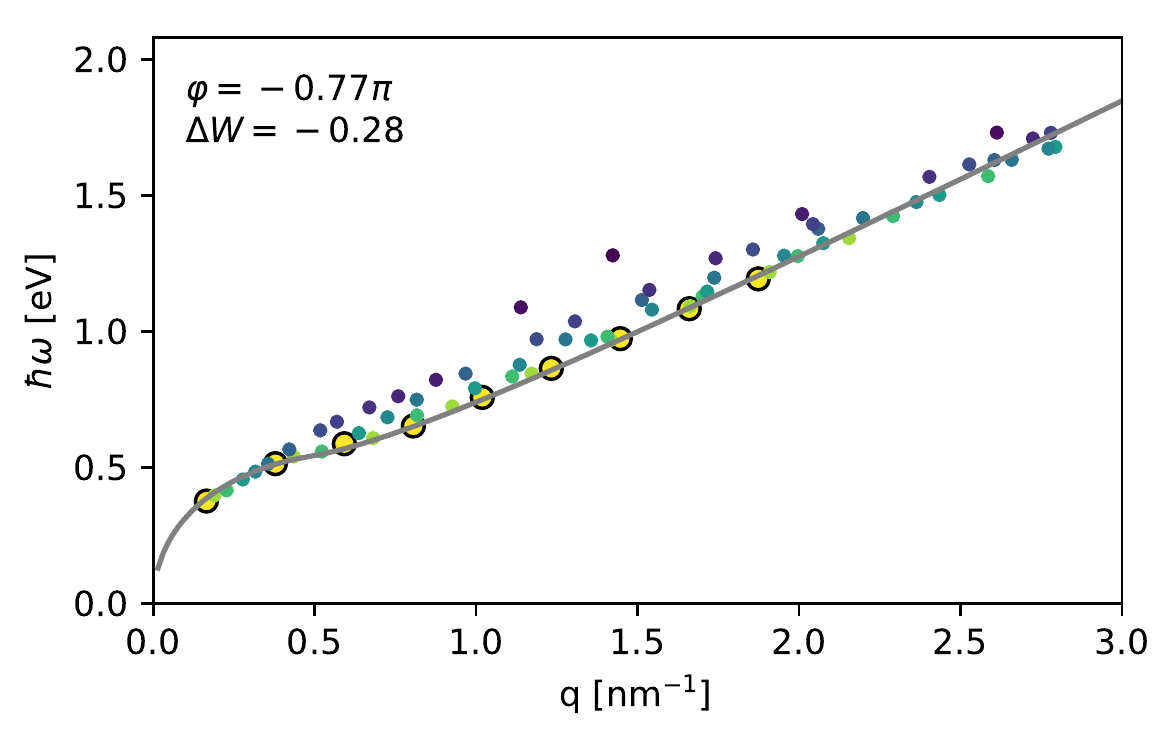}
\caption{(Color online) Optimizing the standing-wave model to the 15 nm zigzag ribbon, the widest ribbon considered here. The model works well for this width, but less so for smaller ribbons, in contrast to what was found for AC ribbons. Colors are the same as in Fig.~\ref{fig:mode_dependency}.}\label{fig:zz_wide_ribbon_phase}
\end{figure}

To get further insight into the plasmons in ZZ ribbons we optimize $\varphi$ and $\Delta W$ for each ribbon width individually.
The results depicted in Fig.~\ref{fig:width_dep_phase} show that there are only minor changes as a function of width for AC ribbons, which is to be expected since one set of (width-independent) parameters did very well previously.
We distinguish between semi-metallic (triangles) and semiconducting AC ribbons and find that they behave slightly different for the small widths, as we have also examined in another context previously.\cite{Wedel2018}
The graphs for the two types of AC ribbons will merge  for wider ribbons (not shown) as the band gap for the semiconducting ribbons closes.

For ZZ ribbons, a standing-wave model with nonlocal bulk dispersion results in much greater variance in the reflection phase and the width correction between the different ribbon widths.
In the zoomed view in the bottom middle panel of Fig.~\ref{fig:width_dep_phase} we can see that only for the two widest $\gtrsim 13$\,nm ribbons (yellow and light green dots) do the TB calculations follow the nonlocal dispersion model well. So it seems that our bulk-dispersion-in-between-reflections model does not apply to the narrower ZZ ribbons that we considered, while for AC ribbons it does for all sizes.

Let us give an explanation why this would be the case.  
The electron density for an AC ribbon is virtually constant across the entire width of the ribbon, see Fig.~\ref{fig:ribbondens}.
Hence, it is a fair assumption that the plasmon experiences a fairly constant bulk-like environment while propagating in between the ribbon edges.
Turning our attention to the electron density in ZZ ribbons, the localized edge states give rise to increased electron density (see second panel of Fig.~\ref{fig:ribbondens}), and therefore an effectively different Fermi energy altering the dispersion of the plasmons in this region.
The effective phase change will thus be the sum of the reflection at the edge and any phase picked up during propagation in the edge region.
With wider ribbons, the relative size of the non-bulk-like region to the plasmon wavelength decreases and the phase shift converges close to $-0.75\pi$ for ZZ ribbons as well.
By comparing to the results from excluding edge states we see that both the phase and the $\Delta W$ vary much less and that the fit hardly changes compared to the width-independent model. The latter was also the case for the AC ribbons.

The ZZ width correction finds its stable point close to -0.3\,nm exactly as the result found for AC ribbons.
Only optimizing for the widest ribbon where the model is applicable yields $\varphi = -0.77\pi$ and the fit shown in Fig.~\ref{fig:zz_wide_ribbon_phase}.

To conclude, a constant phase shift of the same size of $-0.75\pi$ as the ones found in continuum theories works well for both AC and ZZ ribbons, although the picture starts to change for ZZ ribbons narrower than 15\,nm.
At these sizes an atomistic model is needed to properly account for the edge effects.
We must stress that these findings depend on including the width correction, $\Delta W$, not previously considered in earlier work.
Leaving it out yields both different phases and in general worse fits.
Naturally, since $\Delta W$ is on the order of Ångströms, and the plasmon wavelength scales with the ribbon width, its importance will disappear for wide enough ribbons.

\section{Edge-state induced broadening}\label{sec:edgestatesinducedbroadening}

\begin{figure}
\centering
\includegraphics[width=3.4in]{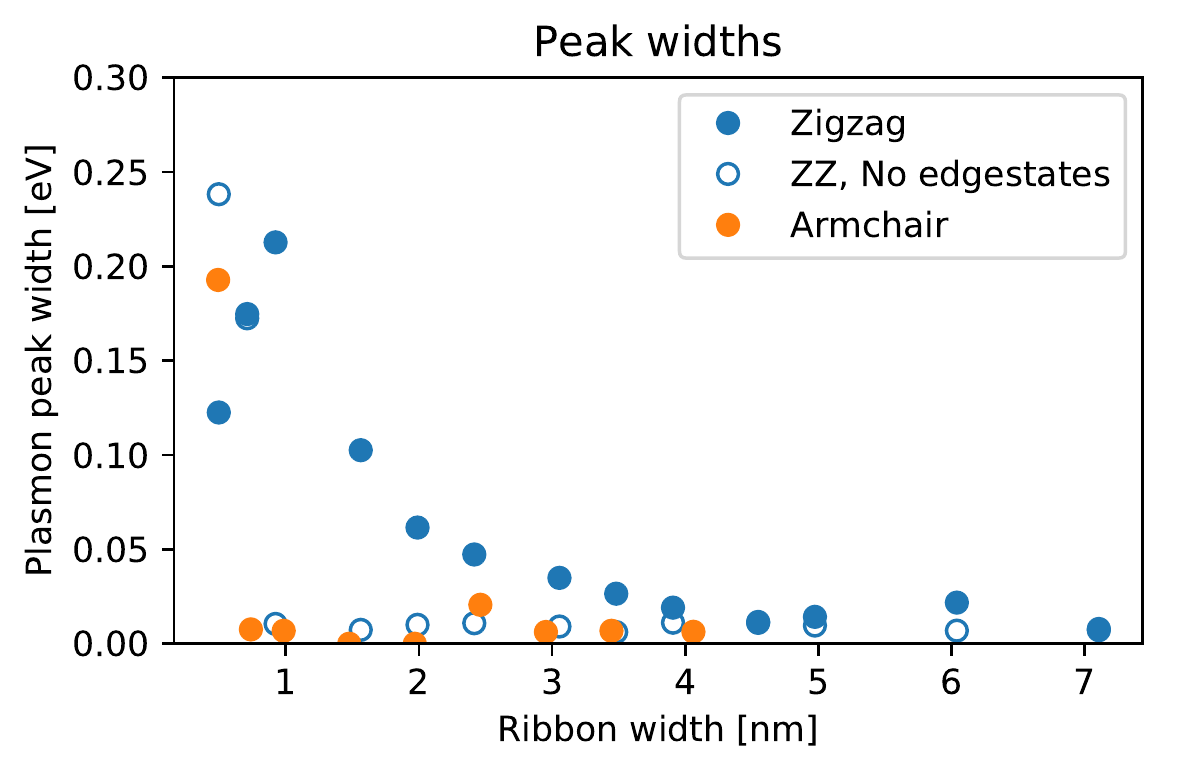}
\caption{(Color online) Width of plasmon peaks when including (full, blue points) and excluding (open, blue points) the edge states from the calculation. Results for armchair ribbons are shown in orange. The edge states contribute with a broadening that increases for smaller widths.}\label{fig:peakwidth}
\end{figure}

Besides the reflection properties dependence on the occurrence of localized edge states we also find that the plasmonic peaks are much wider in ZZ ribbons than in AC ribbons of comparable widths, see Fig.~\ref{fig:peakwidth}.
A similar result has previously been reported in Ref.~\onlinecite{Thongrattanasiri2012QuantumInformation}, and the hypothesis was put forward that the edge states give rise to the additional broadening. Here we will test the hypothesis: by excluding the edge states from the calculation of the optical response, we can directly determine the influence of said states on the broadening. 

The result can be seen 
in Fig.~\ref{fig:peakwidth}, where the blue (orange) dots are the plasmon peak widths for ZZ (AC) ribbons with $\epsilon_F = 0.4\unit{eV}$ and the open symbols are ZZ without edge states. It confirms unequivocally and for the first time the hypothesis that the larger broadening for ZZ ribbons is indeed due to the presence of the edge states. It can be interpreted in this way that the edge states constitute an additional decay channel for the plasmons, leading to more broadening, in an electron energy range that would otherwise have a zero density of states. Indeed, this has been explored analytically for disk resonators\cite{Christensen2014ClassicalStates} and numerically for triangular flakes.\cite{Wang2015PlasmonicNanotriangles}
As edge states are common to all graphene terminations, except the armchair edge\cite{Bellec2014ManipulationGraphene,Akhmerov2008BoundaryLattice,Delplace2011ZakGraphene}, it is reasonable to expect that this edge-induced plasmon broadening will occur in most graphene nano-structures.

\section{Inherited fine structure of plasmonic modes}\label{sec:finestructure}

In this section we will present our findings of the atomic-scale fine structure of the plasmonic modes of nanoribbons.
As the induced charges are built from electron-hole pairs, some structural properties of the underlying wavefunctions will be inherited by the plasmons, as we show in the following.

\subsection{Fine structure of wavefunctions}\label{sec:finestructurewavefunctions}

It is possible to get analytical insight into the shape of the wavefunctions from the Dirac model where the TB Hamiltonian is linearized around the $K$ and $K'$ valleys.
The resulting Hamiltonian has the form 
\begin{align}
H &= \hbar v_\mathrm{F} (\tau_0 \otimes \sigma_x k_x + \tau_z \otimes \sigma_y k_y) \label{eq:hamilton}\\
&=\hbar v_\mathrm{F}\begin{pmatrix}
0 & k_x - ik_y & 0 & 0 \\
k_x + ik_y & 0 & 0 & 0\\
0 & 0 & 0 & -k_x - ik_y\\
0 & 0 & -k_x + ik_y & 0
\end{pmatrix}, \notag
\end{align}
where $\tau_i$ and $\sigma_i$ are all Pauli spin-matrices with the former belonging to valley space and the latter to the $A$/$B$ sub-lattice space.

The armchair edge termination consists of alternating $A$- and $B$-lattice sites and the boundary conditions must thus mix the two valleys\cite{CastroNeto2009}
\begin{align}
\begin{split}
0 &= \phi^{A/B}(x=0) + \phi^{A'/B'}(x=0), \\
0 &= e^{iKW}\phi^{A/B}(x=W) + e^{-iKW}\phi^{A'/B'}(x=W),
\end{split}
\end{align}
where $K = 4\pi/3\sqrt{3}a_0$ and $-K$ are the positions of the $K$-valleys in momentum space and $a_0$ is the interatomic distance in the graphene lattice.
These conditions lead to eigenstates that can be written as a four-vector of plane waves\cite{Brey2006ElectronicEquation}, $e^{ik_nx}$.
We have previously found\cite{Wedel2018} that the allowed values of $k_n$ given in Ref.~\onlinecite{Brey2006ElectronicEquation} can be written in the form
\begin{align}
k_n = \frac{\pi[3n - 2(N+1)]}{3W},
\end{align}
relating the wavelength to three times the width of the ribbon. Here, $N$ is the number of atom rows in the unit cell and $n\in\mathbb{Z}$.
The corresponding eigenenergies are given as $\epsilon = s\hbar v_\mathrm{F}\sqrt{\smash[b]{k_y^2+k_n^2}}$.

\begin{figure*}[htbp]
\centering
\includegraphics[width=7.0in]{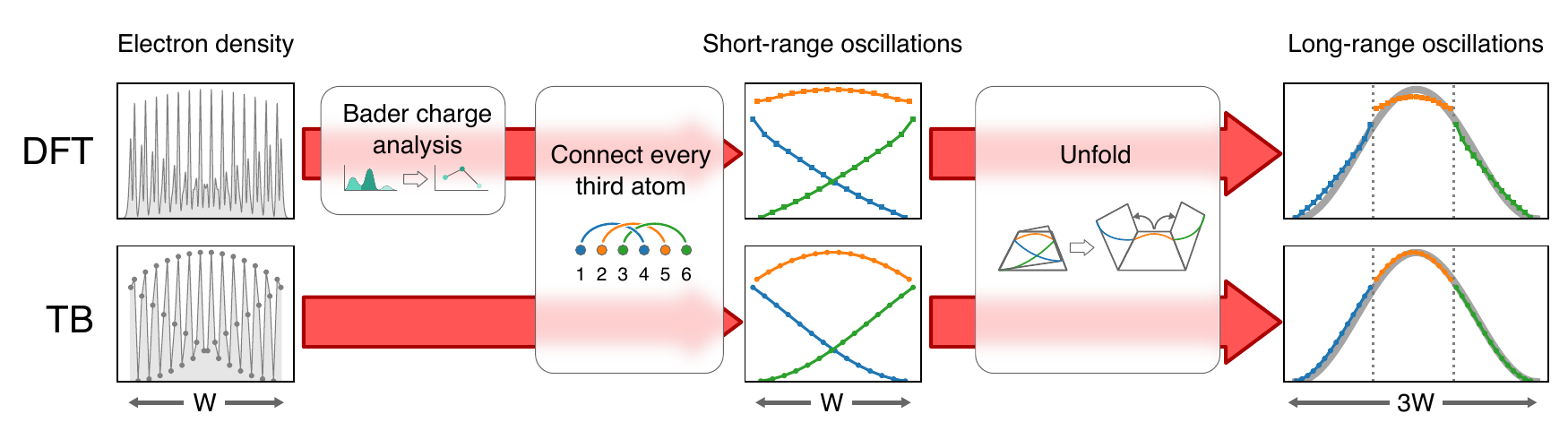}
\caption{(Color online) Scheme for visualizing short- and long-range oscillations in the wavefunctions. Electron density (first column) is mapped to individual atoms and every third atom is connected in the plot (middle column). Finally, the map is ``unfolded'' to reveal the oscillation predicted from the Dirac model.}\label{fig:ac_wf_explained}
\end{figure*}

\begin{figure*}[htbp]
\centering
\includegraphics[width=7in]{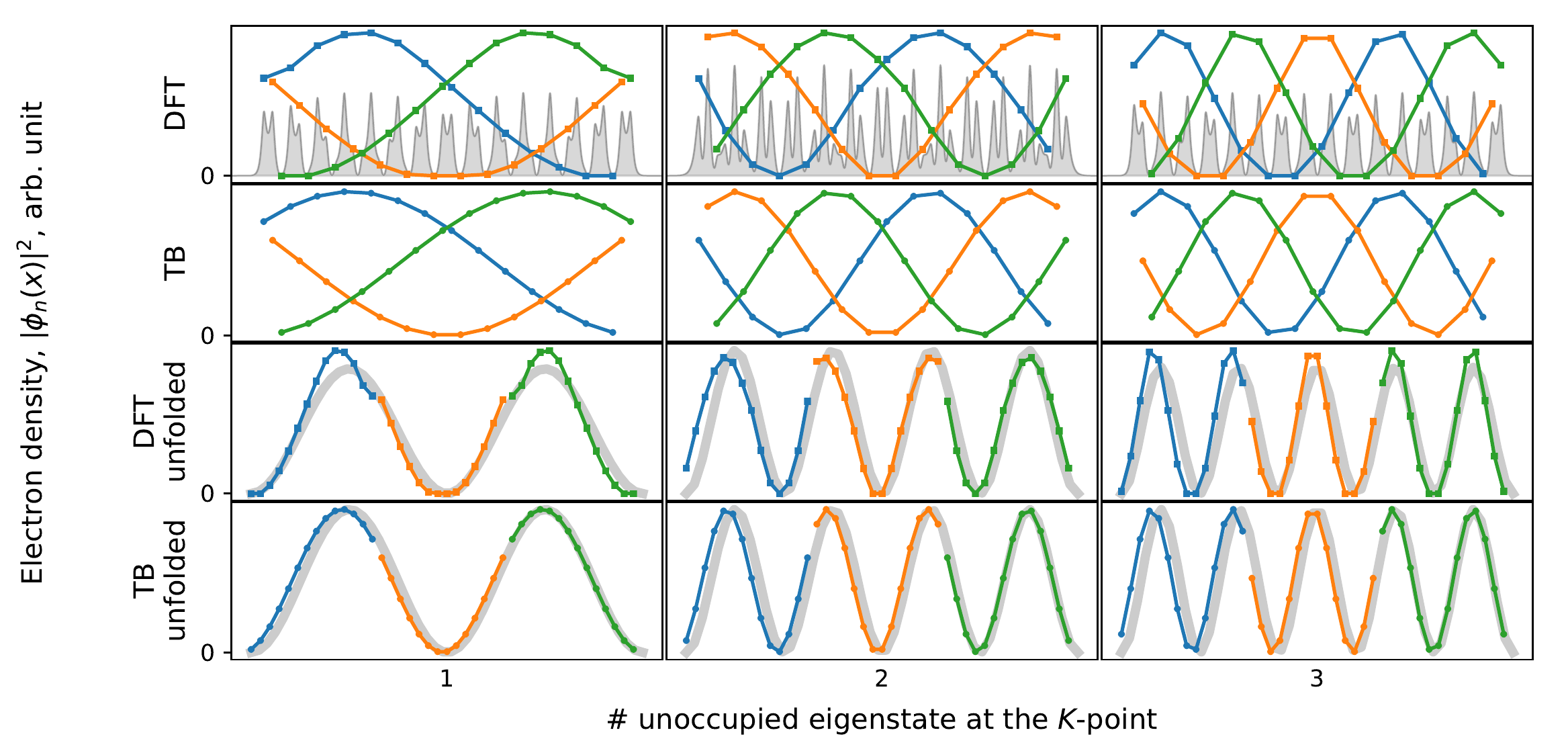}
\caption{(Color online) The electron densities of the three lowest unoccupied wavefunctions at the K-valley. The top row shows the TB results with every third atom connected. The second row shows the DFT electron density (gray) and the result of a Bader charge analysis. The short-wavelength oscillations of every third site are clearly visible. \emph{Bottom row:} Reordering the sites by taking one third at a time reveals the long-wavelength mode. See details in main text.}\label{fig:ac_wf_all}
\end{figure*}

The mixing of the valleys through the boundary conditions will result in an oscillation of the wavefunction\cite{BreyEdgeGraphene} with wavelength $2\pi/K = 3a_0\sqrt{3}/2$ which exactly corresponds to every third atom across the armchair ribbon.
From this it follows that two neighboring atoms will usually have very different weights of the wavefunction. However, if we plot the same electron densities for every third site, such that the atoms 1,4,7,\dots\ are connected, then we expect the change to be rather smooth.
This ``fine structure'' oscillation is readily found in the TB results as shown in Fig.~\ref{fig:ac_wf_explained} and \ref{fig:ac_wf_all} for a 42-atom-wide armchair ribbon.

To emphasize the fundamental nature of this oscillation, we have also performed a DFT calculation of the same ribbon geometry, using  a plane-wave basis set.\footnote{We use the \href{https://wiki.fysik.dtu.dk/gpaw/}{\color{blue}\textsc{GPAW}} code with a cut-off energy of 500 eV and 15 $k$-points in the periodic direction of the supercell.}
Using a Bader charge analysis\cite{Henkelman:2006} we have projected the electron densities corresponding to the lowest unoccupied wavefunctions (of undoped graphene) onto the individual carbon atoms such that we can compare with the TB results.
The \emph{ab initio} calculations show very much the same fine-structure behavior as seen in the top rows of Fig.~\ref{fig:ac_wf_explained} and \ref{fig:ac_wf_all}.

These rapid electronic variations are inherited by the spatial distributions of the \emph{plasmons} of AC graphene ribbons, as we will see in the next section.

Returning to the values of $k_n$ we can also find the long-wavelength oscillation in both the DFT and TB results. As illustrated in Fig.~\ref{fig:ac_wf_explained}, by ``unfolding'' the wavefunction such that it covers the full $3W$, we find that the behavior exactly matches a wave with the shape $\cos(k_nx)$.
It can be seen in Fig.~\ref{fig:ac_wf_all} that this also works for the higher-lying wavefunctions.
Generally, we find that for semiconducting AC ribbons the electron density from state $n$ at site $i$ can be written in as
\begin{align}
\rho_i = \mathcal{N} \sin^2\pb{\pa{x_i- \pb{(i+N)\!\!\!\!\mod 3}}k_n},
\end{align}
where $i$ is the site index as indicated in Fig.~\ref{fig:ac_wf_explained}, $x_i$ is the $x$-coordinate of the site, and $\mathcal{N}$ is a normalization factor.

\subsection{Fine structure of plasmons}\label{sec:finestructureplasmons}

\begin{figure*}[htbp]
\centering
\includegraphics[width=6.5in]{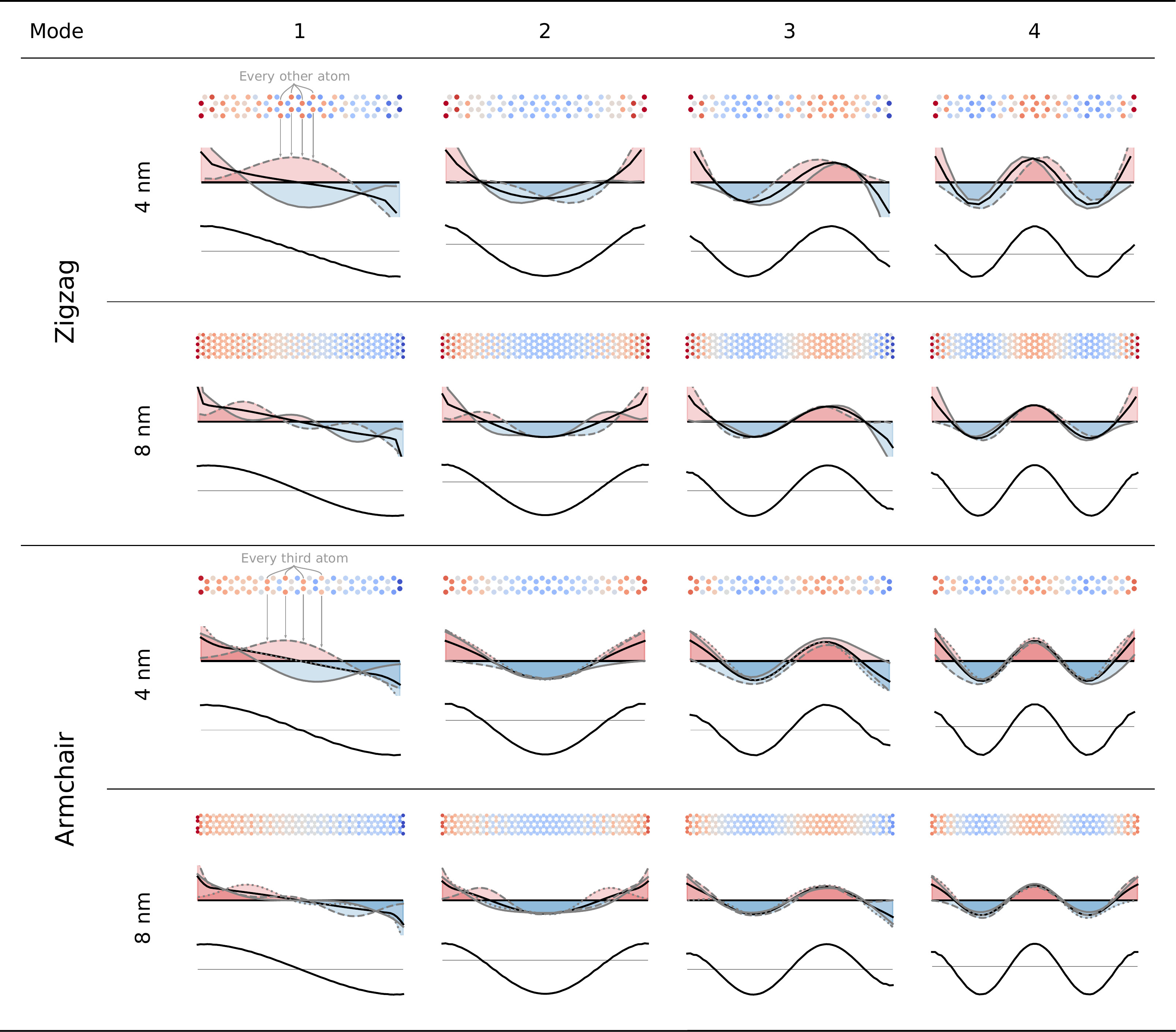}
\caption{(Color online) Induced charges for the first four plasmons for four different ribbons. The top view of the ribbons show the charge on every atomic site. The colored graphs show the charges split between the $A$ and $B$ sublattices for the zigzag ribbon and split between every third atom in the armchair ribbon. The thick line is the average of the thin lines and matches well with the classical expectation. There is a clear fine-structure in the distribution of the charges that seem to disappear at higher-order modes. The bottom graph in each plot shows the induced field. There is evidently considerable fine-structure in the induced charges on the atomic scale.}\label{fig:induced_charges}
\end{figure*}

As explained in the Methods section, the formalism for calculation of the plasmons in TB gives direct access to the induced electron density of the plasmonic modes as well as the induced field through the eigenmodes of the dielectric matrix.
In Fig.~\ref{fig:induced_charges} we show these densities for the four lowest-order modes in two zigzag and armchair ribbons, one $4\unit{nm}$ and one $8\unit{nm}$ of either kind.
For the zigzag ribbon the density is shown on each of the $A$/$B$ sublattices individually (gray lines) as well as the mean density found by averaging two interpolated splines fitted to the sublattice data (thick, black line).
The mean induced density shows the behavior that one would expect in a classical model, but there is a lot of fine-structure oscillations when looking at the atomic details.
The charge fluctuates between the two sublattices, although the variation becomes smaller in the higher-order modes and for the wider ribbons.

Charge densities in the armchair ribbons behave qualitatively different in that there is no $A$/$B$ symmetry as for ZZ.
As explained above, the valley-mixing imposed by the armchair boundary conditions leads to a periodic behavior of the wavefunctions with a characteristic length scale corresponding to every third atom across the ribbon.
We plot the induced charges projected on the three subsets formed by this rule (full, dashed, and dotted gray lines) and find a smooth behavior for all of them.
The fine-structure is thus a fingerprint of the periodicity of the underlying wavefunctions that are involved in building up the plasmon.
As before in Fig.~\ref{fig:ac_wf_all}, in  Fig.~\ref{fig:induced_charges} we show the average induced charges (black lines) and find that they also match very well with the classical picture despite the large local differences.

\section{Discussion and conclusions}\label{sec:conclusions}

Using TB we identify numerous interesting effects in graphene nanoribbon plasmons.
By looking at the dispersion of higher-order plasmons we find edge-dependent reflection properties of narrow ribbons.
For armchair ribbons, the standing waves are well described with a constant phase shift of $-0.75\pi$ and width correction $\Delta W = -0.3\unit{nm}$ at least down to $\sim 2$ nm wide ribbons.
The inclusion of $\Delta W$ is necessary to adequately describe the system within the Fabry--P{\' e}rot model, and leaving it out would render the $-0.75\pi$ phase change inapplicable for the structures considered.
In contrast to the result found for AC ribbons, the $\varphi$ and $\Delta W$ do depend on the width in zigzag ribbons as wide as $\sim 15\unit{nm}$.
This behavior is caused by the localized edge states that significantly alter the electron density close to the ribbon borders.
Surprisingly, at the wider ribbon widths, both ribbon types are characterized with the same width corrections and reflection phases.
These almost identical outcomes were not put in by hand and are the result of independent curve fitting.
So we find that for wide enough ribbons where $\Delta W$ is negligible, the reflection phase of $-0.75\pi$ found in previous numerical studies within continuum models will also work for tight-binding models with either edge termination, a phase which is not far from the value of $-0.64\pi$ found  analytically from a continuum model in  Ref.~\onlinecite{Nikitin2014AnomalousResonators}.
This convergence of our results for the reflection phases of the two ribbon types is consistent with Ref.~\onlinecite{Thongrattanasiri2012QuantumInformation}, where it is shown, using tight-binding calculations,  that in wide ribbons the energies of the lowest-order plasmon of ZZ and AC ribbons coincide.

By looking at the induced charges we find a distinct fine-structure oscillation between the $A$/$B$ sublattice for zigzag ribbon and an every-third atom dependence for the armchair ribbons.
In armchair ribbons, the plasmonic fine-structure oscillations come from similar oscillations in the wavefunctions that are a consequence of the valley-mixing induced by the boundary conditions.
Using analytical results from the Dirac model, we find a general expression for the single-wavefunction electron density around the $K$-point in semiconducting ribbons.

Finally, we have studied edge-induced broadening, which for other geometries was discussed in Refs.~\onlinecite{Christensen2014ClassicalStates,Wang2015PlasmonicNanotriangles}.
We confirmed the hypothesis put forward in Ref.~\onlinecite{Thongrattanasiri2012QuantumInformation} and directly showed the key role played by localized edge states in the broadening of the plasmonic peaks in ZZ ribbons, a broadening that we find is larger for narrower ribbons.
As edge states occur in all but the armchair configuration, we predict that this broadening will be present in most graphene structures.

\section*{Acknowledgments}

This work was supported by the
Danish Council for Independent Research--Natural Sciences (Project 1323-00087). 
The Center for Nanostructured Graphene is sponsored by the Danish National Research Foundation (Project No. DNRF103).
N.~A.~M. is a VILLUM Investigator supported by VILLUM FONDEN (grant No. 16498).
K.~S.~T. acknowledges funding from the European Research Council (ERC) under the European Union's Horizon 2020 research and innovation program (grant agreement No 773122, LIMA).

\bibliography{Mendeley}

\end{document}